\documentclass[pra,twocolumn,showpacs,preprintnumbers,amsmath,amssymb,superscriptaddress]{revtex4}

\usepackage{graphicx}
\usepackage{sidecap}
\usepackage{dcolumn}
\usepackage{bm}

\usepackage[T1]{fontenc}
\usepackage[latin1]{inputenc}
\usepackage[english]{babel}
\usepackage{amsmath}
\usepackage{amssymb}
\usepackage{amstext}
\usepackage{braket}
\usepackage{relsize}
\usepackage{subfigure}

\newcommand{\ud}{\mathrm{d}}
\newcommand{\ue}{\mathrm{e}}

\newcommand{\vect}[1]{\ensuremath{\bm{#1}}}

\begin{document}

\preprint{}

\title{Orbital order of spinless fermions near an optical Feshbach resonance
}

\author{Philipp Hauke}
	\email{philipp.hauke@icfo.es}
	\affiliation{ICFO -- Institut de Ci\`{e}ncies Fot\`{o}niques, Parc Mediterrani de la Tecnologia, 08860 Castelldefels, Spain}
	\affiliation{Kavli Institute for Theoretical Physics,
	University of California, Santa Barbara, CA 93106}
\author{Erhai Zhao}
	\affiliation{Department of Physics and Astronomy, George Mason University, Fairfax, VA 22030}
	\affiliation{Kavli Institute for Theoretical Physics, University of California, Santa Barbara, CA 93106}	
\author{Krittika Goyal\footnote{K.~G.\ previously published under the name K.\ Kanjilal.}}
	\affiliation{Center for Quantum Information and Control (CQuIC), and Department of Physics and Astronomy, University of New Mexico, Albuquerque NM 87131}	
\author{Ivan H.\ Deutsch}
	\affiliation{Center for Quantum Information and Control (CQuIC), and Department of Physics and Astronomy, University of New Mexico, Albuquerque NM 87131}	
\author{W.\ Vincent Liu}
	\affiliation{Department of Physics and Astronomy, University of Pittsburgh, Pittsburgh, PA 15260}
	\affiliation{Kavli Institute for Theoretical Physics, University of California, Santa Barbara, CA 93106}	
\author{Maciej Lewenstein }
	\affiliation{ICFO -- Institut de Ci\`{e}ncies Fot\`{o}niques, Parc Mediterrani de la Tecnologia, 08860 Castelldefels, Spain}
	\affiliation{ICREA -- Instituci{\'o} Catalana de Recerca i Estudis Avan\c{c}ats, Lluis Companys 23, E-08010 Barcelona, Spain}
	\affiliation{Kavli Institute for Theoretical Physics, University of California, Santa Barbara, CA 93106}

\date{\today}

\begin{abstract}
We study the quantum phases of a three-color Hubbard model that arises in the dynamics of the $p$-band orbitals of spinless fermions in an optical lattice.  Strong, color-dependent interactions are induced by an optical Feshbach resonance. Starting from the microscopic scattering properties of ultracold atoms, we derive the orbital exchange constants at 1/3 filling on the cubic optical lattice.  
Using this, we compute the phase diagram in a Gutzwiller ansatz. 
We find novel phases with `axial orbital order' in which $p_z$ and $p_x + i p_y$ (or $p_x - i p_y$) orbitals alternate.
\end{abstract}


\maketitle
 
Orbital physics of electrons plays an important role in strongly-correlated solid-state systems, \textit{e.g.}, transition-metal oxides (see, \textit{e.g.}, \cite{Tokura2000,Horsch2007} and references therein). 
In particular, intriguing quantum phases emerge due to the coupling of the orbital degree of freedom to the charge, spin, or lattice degrees of freedom \cite{Kugel1982,Khaliullin2005}. 
Such coupling, while leading to interesting effects, also complicates the theoretical treatment. It is, therefore, desirable to study simpler systems with the orbital degree of freedom decoupled from all others. 
Ultracold atoms in higher bands of optical lattices provide an ideal tool to study orbital dynamics in a well-controlled environment, including orbital-only models of single-species (spinless) fermions.

Several groups have now achieved loading and manipulating ultracold atoms in higher (such as $p$-) bands of optical lattices 
\cite{Browaeys:05,Esslinger:05,MuellerBloch:07,Anderlini:07,Wirth2011}.
Techniques such as lattice ramping or radio-frequency pulses have been used to transfer atoms from the $s$- to higher bands, 
where they can stay in a metastable state for a sufficiently long time.
For spinless fermionic atoms, the $p$-band can also be simply populated by first completely filling the $s$-band, 
requiring larger particle numbers, but less experimental control.
To avoid undesired collisions between ground and excited-band atoms, the $s$-band atoms may be removed afterwards using laser pulses \cite{footnoteCollisionsppsd}.

The interaction between fermionic atoms is usually weak at low temperatures because the Pauli exclusion principle 
only allows scattering in high partial-wave channels ($p$, $f$, {\it etc.}). 
One way to increase the $p$-wave elastic scattering cross section is to employ a Feshbach resonance (FR) \cite{Regal2003}. 
Typically, this is done by coupling channels in the electronic ground state through magnetic fields.  
For the case of $p$-waves, however, this method usually leads to significant atom losses through three-body inelastic collisions 
because the scattering state is well localized by the angular momentum barrier, and has good Franck-Condon overlap with more deeply bound molecules \cite{chin-rev}. 
To circumvent this problem, recently Ref.~\cite{Goyal2010b} considered enhanced $p$-wave interactions via an optical FR (OFR) between a scattering state
 and an electronically excited ``purely-long-range'' molecule.  
Such molecules have inner turning points at very large distances (\emph{e.g.}, $>50 a_0$ in $^{171}$Yb), 
well beyond the chemical binding region, and thus three-body recombination should be highly suppressed. 
This approach not only allows to study strongly-correlated phases, but also provides for a high degree of control. 
In particular, the interaction strength among different $p$-orbitals can be tuned differently.

Motivated by these developments, we investigate in this article the phase diagram of 
spinless fermions on a cubic lattice near an OFR described by the following Hubbard-like model: 
\begin{eqnarray}
\label{eq:H}
H=-\sum_{\vect{i}; \mu,\nu}
t_{\mu,\nu} (c_{\mu,\vect{i}}^\dagger c_{\mu,\vect{i}+\vect{e}_{\nu}} + h.c.) 
+\sum_{\vect{i}} \big[ V_1 n_{x,\vect{i}} n_{y,\vect{i}}  \nonumber \\
+  V_2 ( n_{x,\vect{i}} n_{z,\vect{i}} + n_{y,\vect{i}} n_{z,\vect{i}} ) 
+ ( i V_3 c_{x,\vect{i}}^\dagger c_{y,\vect{i}} n_{z,\vect{i}} + h.c.) \big]. 
\end{eqnarray}
The operator $c_{\mu,\vect{i}}$ destroys a fermion in the orbital $p_{\mu}$ 
at site $\vect{i}$, and $n_{\mu,\vect{i}}$ is the corresponding number operator.
The lattice spacing is set to 1, $\vect{e}_{\nu}$ is the unit vector in direction $\nu$, 
and $\mu,\nu=x,y,z$.
The nearest-neighbor hopping amplitude $t_{\mu,\nu}$ describes hopping of fermions in orbital $p_{\mu}$ along the direction $\vect{e}_{\nu}$.
Due to the anisotropy of the $p$-orbital Wannier wave functions, it is direction and orbital dependent
\cite{Girvin:05,Kuklov:06,Liu-Wu:06}, $t_{\mu,\nu}=t_\parallel \delta_{\mu,\nu} + t_\perp \left(1-\delta_{\mu,\nu}\right)$.

The interactions $V_{1,2,3}$ are induced by an OFR laser \cite{Goyal2010b} 
which couples the electronic ground state of the atom to an excited state. 
The interaction can be expressed in terms of the ($p$-wave) pseudo-potential 
$V_p^{m}$ for two particles with mass $M$ and relative angular momentum $m$,
$V_p^{m}\left(\vect{r}\right)=\lim_{s\to 0}\frac{3R}{2M}
\frac{\delta\left(r-s\right)}{s^3}\partial_r^3 r^2$.
The real part of the $p$-wave scattering volume, $R=\mathrm{Re} [\left(a_p^m\right)^3]$, can be tuned by the detuning and the intensity of the OFR laser.
Expanding field operators in the Wannier basis,
$\psi\left(\vect{r}\right)=\sum_{\vect{i},\mu} w_{\mu}\left(\vect{r}-\vect{i}\right) c_{\mu,\vect{i}}$,
the interaction term $\int\ud^3r_1\int\ud^3r_2 \psi^\dagger(\vect{r}_1)\psi^\dagger(\vect{r}_2)V_p^{m}(\vect{r}_1-\vect{r}_2) \psi(\vect{r}_1)\psi(\vect{r}_2)$ leads to the on-site, inter-orbital interaction
$H_{\mathrm{int}}=\sum_{\vect{i}} V_{\mu,\nu,\mu^\prime,\nu^\prime} c_{\mu^\prime,\vect{i}}^\dagger c_{\nu^\prime,\vect{i}}^\dagger c_{\mu,\vect{i}} c_{\nu,\vect{i}}
$, 
where repeated indices are summed over. (We neglect all off-site interactions.)
The matrix element 
$V_{\mu,\nu,\mu^\prime,\nu^\prime}=\sum_{m} \int\ud^3r_1\int\ud^3r_2 w_{\mu^\prime}\left(\vect{r}_1-\vect{i}\right) w_{\nu^\prime}\left(\vect{r}_2-\vect{i}\right) V_p^{m}\left(\vect{r}_1-\vect{r}_2\right)\\ w_{\mu}\left(\vect{r}_1-\vect{i}\right) w_{\nu}\left(\vect{r}_2-\vect{i}\right)$ can now be computed 
by separating the relative and center-of-mass coordinates. 
For deep lattices, the $p$-orbital Wannier functions are well approximated 
by the first excited states of harmonic oscillators 
(with the oscillator length $\zeta$ controlled by the lattice depth). 
The only non-zero interaction terms are the ones given in Eq.~\eqref{eq:H}, with 
$V_1=\frac{1}{4}\left(U_{1}+U_{-1}\right)$, $V_2=\frac{1}{8}\left(U_{1}+U_{-1}+2U_0\right)$, and $V_3=\frac{1}{8}\left(U_{-1}-U_{1}\right)$. 
Here, $U_{m}={3\sqrt{2}R}/({\sqrt{\pi}\zeta^5M})$ defines the interaction strength in the scattering channel 
with angular momentum $m=1,0,-1$. 
A Zeemann splitting, which may be introduced by a magnetic field, 
leads to different detuning of the OFR laser for the three scattering channels. This 
makes the scattering length $a_p^m$ dependent on $m$, and consequently the $U_{m}$'s can be
different in magnitude and even in sign. 
Thus, the relative strengths and signs of $V_{1,2,3}$ can be varied by 
changing the strength of the Zeemann splitting together with the detuning of the OFR laser.
By contrast, in a standard magnetic FR, $U_{-1}=U_{+1}$.
In our case, breaking the symmetry between $U_{-1}$ and $U_{+1}$ leads to the orbital-changing term $V_3$.
Physically, it allows ($p_x$ or $p_y$) particles to move on the two dimensional plane, instead
of along a chain only.
Since it explicitly breaks time-reversal symmetry (TRS), we can expect it to lead to novel phases reflecting that intriguing property.

Hamiltonian \eqref{eq:H} generalizes the models of Refs.\ \cite{Rapp2007,Rapp2008,Zhao2008,wu08,Miyatake2009,Toth2010}.
For $V_1=V_2$, and $V_3=0$, it reduces to the SU(3) Hubbard model.
One can visualize $p$-band fermions as particles carrying a color index 
representing the $p_x$, $p_y$, and $p_z$ orbital states. Then, Hamiltonian \eqref{eq:H} 
describes a three-color fermion model with color-dependent interaction, a novel color-changing term $V_3$, and spatially anisotropic and color-dependent tunneling. We will show below that this model has a rich phase diagram with novel phases. 
Here, we focus on the strong-coupling limit for $p$-band filling 1/3, and determine
the orbital order using a Gutzwiller mean-field ansatz.

\begin{figure}
\includegraphics[width=0.4\textwidth]{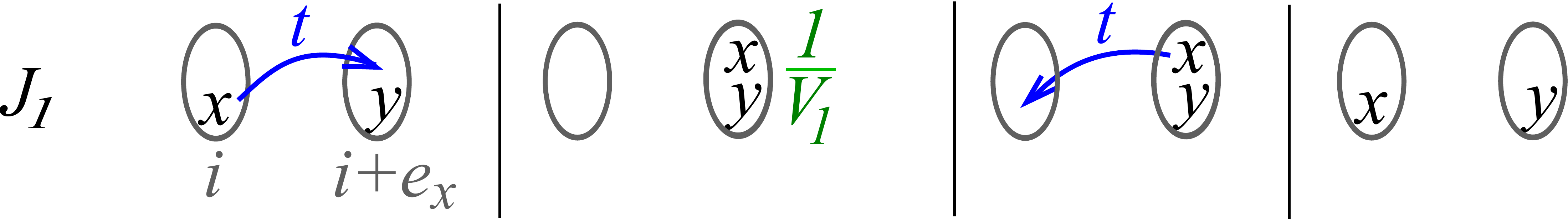}\\
\vspace*{0.6cm}
\includegraphics[width=0.4\textwidth]{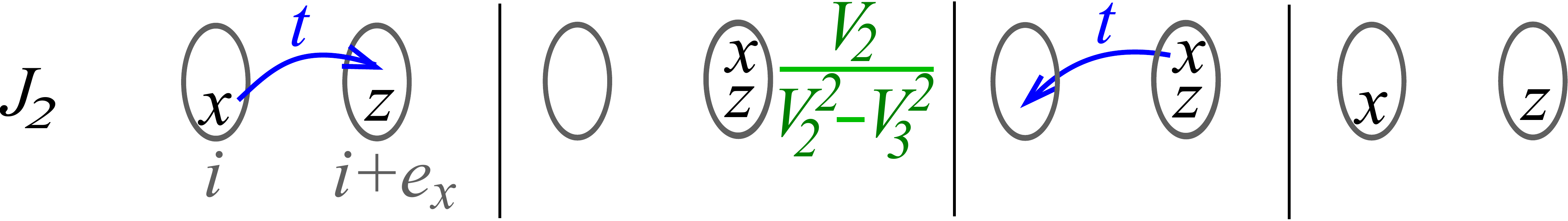}\\
\vspace*{0.6cm}
\includegraphics[width=0.4\textwidth]{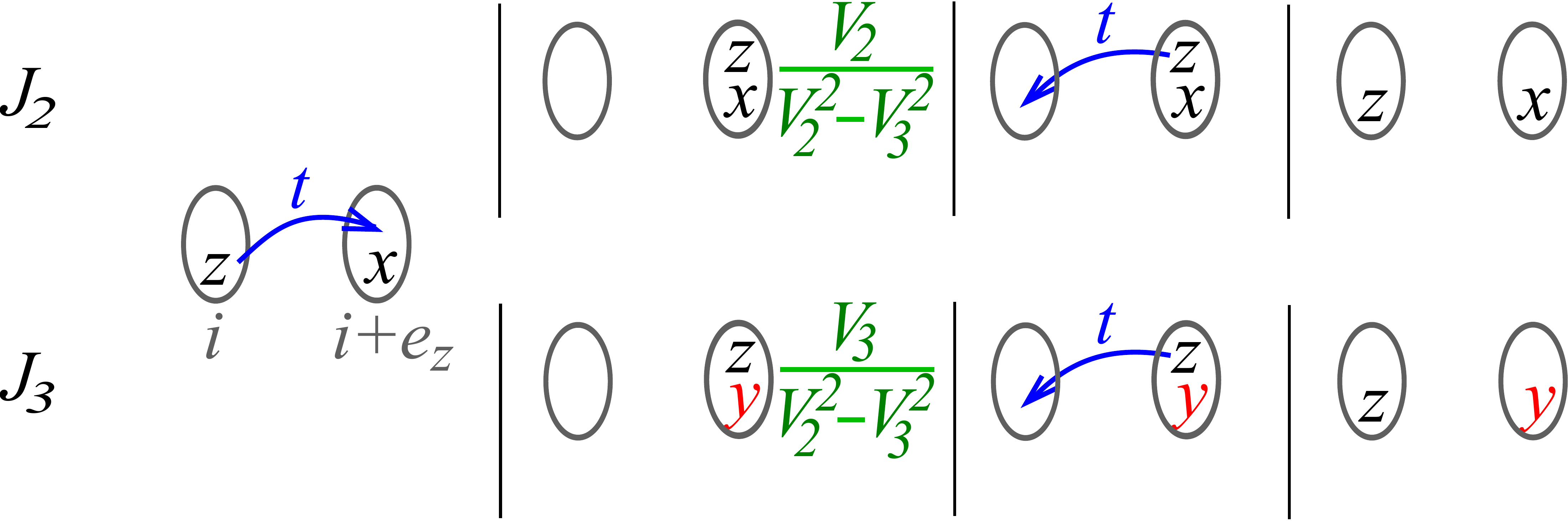}
\caption{
Sketch of the virtual hopping processes leading to the effective Hamiltonian \eqref{eq:Heffonethird}. 
Neglecting $t_\perp$, these -- plus the ones obtained by interchanging $x$ and $y$ -- are the only ones.
Note in particular the orbital-changing process $J_3$. 
Gray ovals denote sites, the blue $t$ tunneling processes, and the green fractions denote interactions. Orbitals $p_{\mu}$ are abbreviated as $\mu$.
\label{fig:virtualHopping}
}
\end{figure}

In the strong-coupling limit,
\begin{equation}
\label{eq:Vggt}
\left|t_\parallel\right|\ll V_1 ,\quad \left|t_\parallel\right|\ll V_2-V_3 ,\quad \mathrm{and}\quad \left|t_\parallel\right|\ll V_2+V_3\,,
\end{equation} 
double occupancy of the same site is suppressed. At $1/3$ filling of the $p$-band, 
there is on average one $p$-band particle per site, and density 
fluctuations are frozen. Virtual hopping 
induces exchange interactions between nearest-neighbor orbitals (see Fig.\ \ref{fig:virtualHopping}). 
The situation bears some resemblance to the emergence of magnetic models, such as the Heisenberg model,
in the strong-coupling limit of the Hubbard model. The difference here is that
three orbital (instead of two spin) states are involved. 
Since $\left|t_\perp\right|\ll \left|t_\parallel\right|$, perpendicular tunneling $t_\perp$ can 
safely be neglected \cite{Zhao2008},
and, for brevity, we write $t=t_\parallel$.
Treating the tunneling $t$ in \eqref{eq:H} as a perturbation and following 
standard second-order perturbation theory, we obtain the effective Hamiltonian
for $1/3$ filling
\begin{align}
\label{eq:Heffonethird}
H_{\mathrm{eff}}=-\sum_{\vect{i}} \Big[ &\sum_{\mu=x,y,z} \sum_{\vect{\delta}=\pm\vect{e}_{\mu}} J_{\mu} n_{\mu,\vect{i}}\left(1-n_{\mu,\vect{i}+\vect{\delta}}\right)  \nonumber \\
					+ &\sum_{\mu=x,y} \sum_{\vect{\delta}=\pm\vect{e}_{\mu}} \left(J_2-J_1\right) n_{\mu,\vect{i}}n_{z,\vect{i}+\vect{\delta}} \nonumber \\
					- &\sum_{\vect{\delta}=\pm\vect{e}_z} J_3( i c_{x,\vect{i}}^\dagger c_{y,\vect{i}} n_{z,\vect{i}+\vect{\delta}}  +h.c. )\Big] \,,
\end{align}
where we have used the constraint $n_{x,\vect{i}}+n_{y,\vect{i}}+n_{z,\vect{i}}=1$, and defined
$J_1\equiv t^2/V_1$, $J_2\equiv t^2 V_2 / (V_2^2-V_3^2)$, $J_3\equiv t^2 V_3/(V_2^2-V_3^2)$, and 
$J_{x}=J_{y}=J_1$, $J_{z}=J_2$.
For $V_3=0$, $V_1=V_2$, Eq.\ \eqref{eq:Heffonethird} reduces 
to $J_{\mu} n_{\mu,\vect{i}}n_{\mu,\vect{i}+\vect{\delta}}$, 
a hallmark of the quantum 3-state Potts-like model \cite{note}.

To see which orbital order is favored, 
we first discuss the simple case of $J_3=0$.
The first term of Eq.\ \eqref{eq:Heffonethird} always favors configurations where the orbitals at neighboring sites differ. 
(A) For $J_1>\max\left(J_2,0\right)$, both the first and second terms favor an alternating pattern between $p_x$- and $p_y$-particles in the $xy$-plane.
(B) For $J_2>\max\left(J_1,0\right)$, the favored configuration is an alternating pattern between $p_z$ and not-$p_z$. 
(C) For (the unstable case) $J_1,J_2<0$, the best configuration is a homogeneously filled lattice. 

Certain aspects of Hamiltonian \eqref{eq:Heffonethird} become clearer when we rewrite 
it in terms of the generators of the SU(3) group. 
In terms of the Gell-Mann matrices $\lambda^{(i)}$ and the so-called $F$-spin operators
$Y=\frac{1}{\sqrt{3}} c_\mu^\dagger \lambda^{(8)}_{\mu, \nu} c_\nu$ and $T^{(\alpha)}=\frac{1}{2} c_\mu^\dagger \lambda^{(\alpha)}_{\mu, \nu} c_\nu$ ($\alpha=1,2,3$), $H_{\mathrm{eff}}$ becomes
\begin{align}
\label{eq:HeffonethirdSU3}
H_{\mathrm{eff}}&= \frac{4}{3} \sum_{\vect{i}} \left[ \left( J_2- J_1 \right)Y_{\vect{i}} - J_3 T^{(2)}_{\vect{i}} \right] \nonumber \\
&+ 2\sum_{\vect{i}} \Big[\sum_{\vect{\delta}=\vect{e}_x,\vect{e}_y} \big(
 J_1 T^{(3)}_{\vect{i}} T^{(3)}_{\vect{i}+\vect{\delta}} + \frac{2J_2-J_1}{4} Y_{\vect{i}} Y_{\vect{i}+\vect{\delta}}   \nonumber \\
&\qquad\qquad\qquad\quad +  \frac{J_2}{2}T^{(3)}_{\vect{i}}Y_{\vect{i}+\vect{\delta}} + \frac{J_2}{2}Y_{\vect{i}}T^{(3)}_{\vect{i}+\vect{\delta}} \big)  \\						
&+  J_2Y_{\vect{i}}Y_{\vect{i}+\vect{e}_z} + J_3T^{(2)}_{\vect{i}}Y_{\vect{i}+\vect{e}_z} +J_3Y_{\vect{i}} T^{(2)}_{\vect{i}+\vect{e}_z}  \Big]  \nonumber  \,,
\end{align}
where we neglected constant terms.
In the basis $\left(p_x,p_y,p_z\right)$, $Y$ and $T^{(3)}$ are diagonal, which means that terms such as $Y_{\vect{i}} Y_{\vect{j}}$, $Y_{\vect{i}} T^{(3)}_{\vect{j}}$, or $T^{(3)}_{\vect{i}} T^{(3)}_{\vect{j}}$ are Ising-like. 
The orbital-changing term $V_3$ leads to $T^{(2)}=\frac{1}{2i}\left(T^{(+)}-T^{(-)}\right)$, where $T^{(\pm)}$ are ladder operators of the $T$-spin.
$T^{(3)}$ and $T^{(2)}$ do not commute, but both commute with $Y$. 
This means that one can replace $Y$ by its eigenvalues $-\frac{2}{3}$
(for $\ket{p_z}$) and $\frac{1}{3}$ (for $\ket{p_x}$ and $\ket{p_y}$),
which gives some insight into the physics of Hamiltonian
\eqref{eq:HeffonethirdSU3}. 
Assuming that the ground state is bipartite with respect to the eigenvalue of $Y$ \cite{inPrincipleNonBibartite},
there are three different cases:
(A) at all sites the eigenvalue of $Y$ is $\frac{1}{3}$, (B) the eigenvalues $-\frac{2}{3}$ and $\frac{1}{3}$ alternate, and (C) all sites have eigenvalue $-\frac{2}{3}$.
In the last case, there is one $\ket{p_z}$-particle per site, whence there is no virtual tunneling, and the Hamiltonian vanishes. 
In the sectors A and B, it reads (neglecting constant terms)
\begin{subequations}
\begin{eqnarray}
	H_{\mathrm{eff}}^{(A)}  &=& \frac{J_1}{2} \sum_{\vect{i}} \sum_{\vect{\delta}=\vect{e}_x,\vect{e}_y} \sigma_{\vect{i}}^{(3)} \sigma_{\vect{i}+\vect{\delta}}^{(3)}  \,; \\
	H_{\mathrm{eff}}^{(B)} &=& -2 J_3 \sum_{\vect{i}\in\Omega} \sigma_{\vect{i}}^{(2)}\,.
\end{eqnarray}
\end{subequations}
Here, $\sigma$ denotes the usual Pauli matrices, which act on the subspace spanned by $\ket{p_x}$ and $\ket{p_y}$.
Sector A is reduced to the Ising model on decoupled $xy$-planes, which favors an antiferromagnetic ground state. This is just the model found in the 2D-case treated in \cite{Zhao2008,wu08}.
In sector B, $\Omega$ denotes the partition where $Y$ has 
eigenvalue $\frac{1}{3}$. On these sites, $J_3$ acts as a magnetic
field in the $y$-direction, lifting the degeneracy between $\ket{p_x}$ and
$\ket{p_y}$ and leading to the ground state
$\left(\ket{p_x}\pm i \ket{p_y}\right)/\sqrt{2}$ (for $J_3\gtrless 0$).

Having obtained a qualitative picture of the expected phases, we now
analyze the phase diagram of Hamiltonian \eqref{eq:HeffonethirdSU3}
quantitatively.  
To this, we assume that correlations between sites
are small so that the ground state can be approximated by a product over sites.  
To find the ground state of Hamiltonian \eqref{eq:HeffonethirdSU3}, we employ the Gutzwiller variational wave
function $\ket{\Psi}=\bigotimes_i (\cos\theta\ket{p_x}_i+\sin\theta\cos\phi\ket{p_y}_i+\sin\theta\sin\phi\ket{p_z}_i)$, which is a product over sites $i$, and minimize the
energy of a cube with side length $L$ (up to $L=8$) under periodic boundary conditions.  
Note, however, that close to phase transitions, where fluctuations become important, such a mean-field ansatz is not valid.

\begin{figure}
\centering
\includegraphics[width=0.5\textwidth]{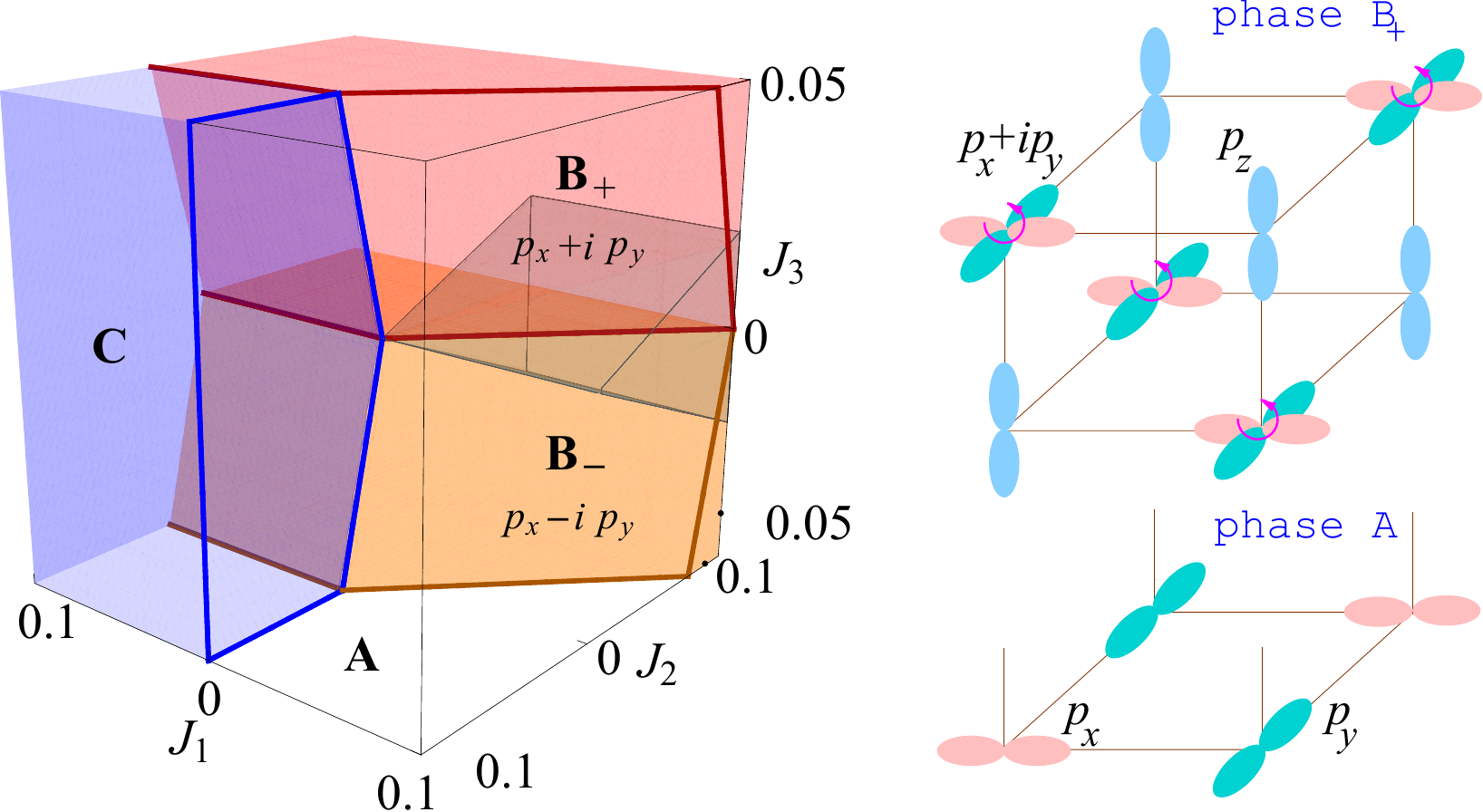}
\caption{ 
{\bf{Left:}}
The phase diagram of $H_{\mathrm{eff}}$ [Eq.~\eqref{eq:Heffonethird}] at $1/3$ filling
shows four phases: (A) antiferro-orbital order (empty region), (B$_+$) axial orbital order
(red region, $J_3>0$) and similarly (B$_-$) (orange region, $J_3<0$), and 
finally (C) with tunneling completely frozen (blue region). 
The gray wedge indicates the region satisfying the strong-coupling conditions \eqref{eq:Vggt}, $0\leq J_{1,2}\ll 1$, $J_3\ll J_2$.
{\bf{Right:}} sketch of phase
B$_+$, in which $\ket{p_z}$ and $\ket{p_x}+ i \ket{p_y}$ orbitals alternate, and phase A.
Phase B$_-$ can be visualized from phase B$_+$ by replacing $\ket{p_x}+ i \ket{p_y}$ with 
$\ket{p_x}- i \ket{p_y}$.
\label{fig:phdonethirdfilling}
}
\end{figure}

The energy per site for even $L$ is smaller than for odd
$L$, showing that the ground state periodicity is indeed 2
\cite{checkLeven}.
In agreement with the qualitative picture above, we find three classes of ground states with different orbital order (summarized in Fig.\ \ref{fig:phdonethirdfilling}):
(A) For $J_1>J_2+\left|J_3\right|/2$ and $J_1>0$ we find an `antiferromagnetic phase' similar to, \textit{e.g.}, the 2D-model of Ref.\ \cite{Zhao2008}: in each $xy$-plane, sites with $p_x$- and $p_y$-orbitals alternate (similar to the antiferromagnetic N\'{e}el state). 	
	Since $p_x$- and $p_y$-particles do not tunnel in the $z$-direction, the $xy$-planes are decoupled, and within our approximation (\emph{e.g.}, neglecting $t_\perp$), 
	there is no long-range order in the $z$-direction. 
	It is possible, however, that long-range order among the planes develops at low temperature for finite $t_{\bot}$. 
(B) For $J_1<J_2+\left|J_3\right|/2$ and $J_2>-\left|J_3\right|/2$ the ground state shows \emph{axial orbital order}. 
	The state is bipartite with $\ket{p_z}$ on one sublattice and $\left(\ket{p_x}\pm i \ket{p_y}\right)/\sqrt{2}$ (for $J_3\gtrless 0$, respectively) 
	on the other sublattice (right-hand panel of Fig.\ \ref{fig:phdonethirdfilling}). The degeneracy between $\ket{p_x}$ and $\ket{p_y}$ is lifted by a finite $J_3$. 
	The state $\left(\ket{p_x}\pm i \ket{p_y}\right)/\sqrt{2}$ has \emph{finite angular momentum}, this novel phase breaks TRS \cite{footnoteOffsiteinteractions}. 
(C) For $J_1<0$ and $J_2<-\left|J_3\right|/2$ Pauli exclusion prohibits all tunneling $t_\parallel$ (by filling $\alpha\beta$-planes ($\alpha\beta=xy,xz,yz$) uniformly with $p_{\alpha}$ or $p_{\beta}$). 
	This state is unstable, however, because it 
	cannot fulfill the strong-coupling requirements \eqref{eq:Vggt}.
	Interestingly, phases A and C preserve TRS, although $V_3$ in Hamiltonian \eqref{eq:H} breaks it explicitly.

\begin{figure}
\vspace*{0.3cm}
\includegraphics[width=0.14\textwidth]{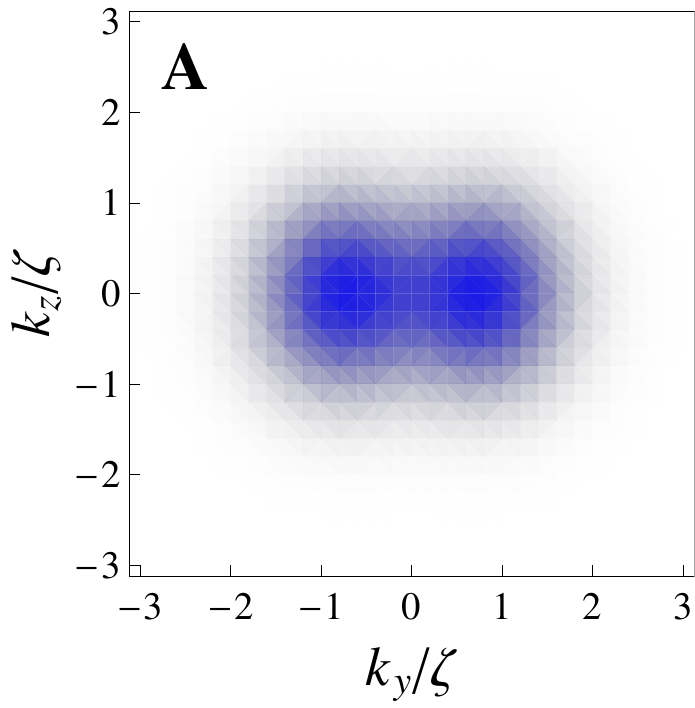}\quad\includegraphics[width=0.14\textwidth]{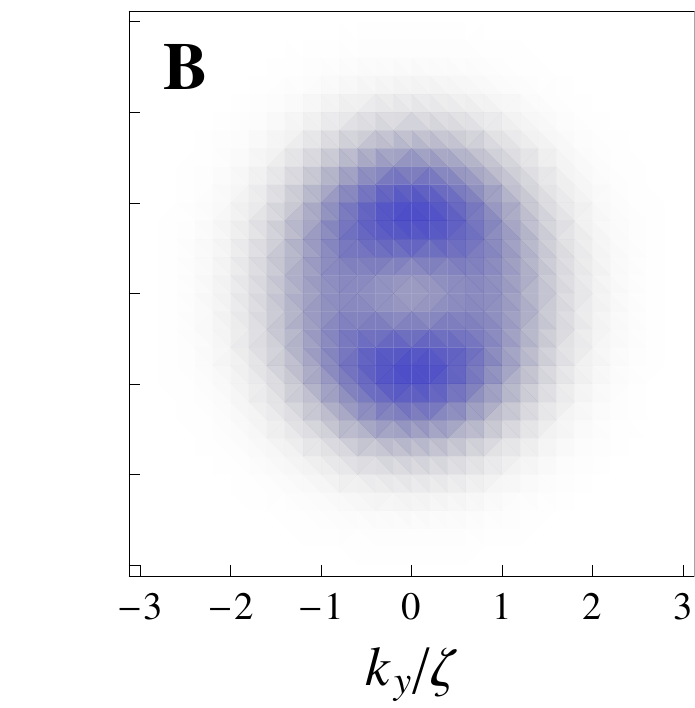}\\
\vspace*{0.15cm}
\includegraphics[width=0.14\textwidth]{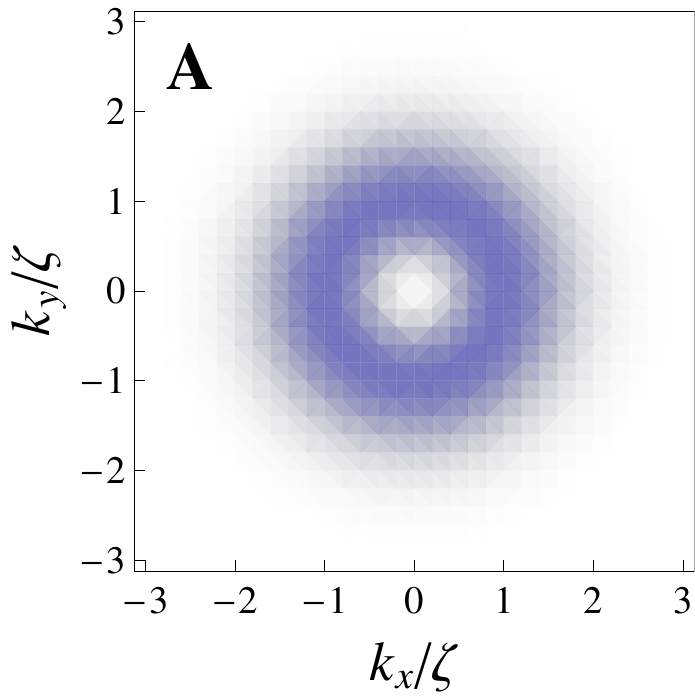}\quad\includegraphics[width=0.14\textwidth]{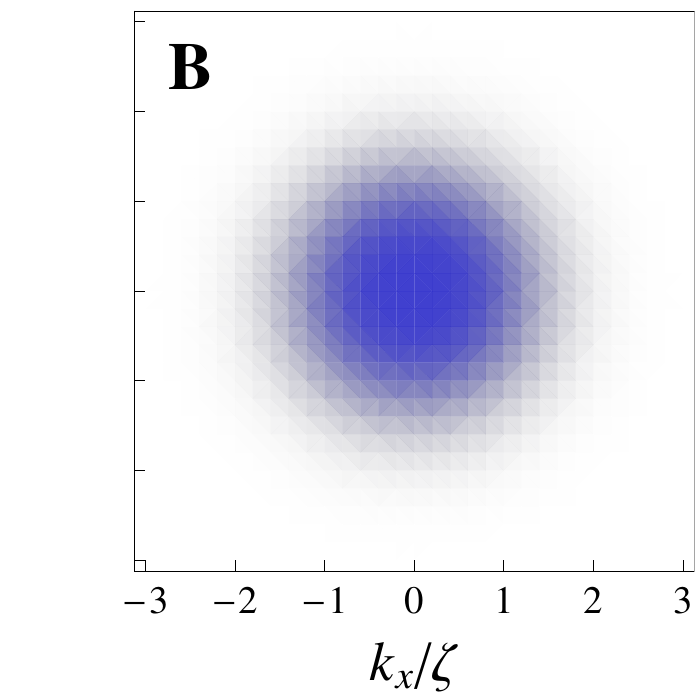}
\caption{
Predicted time-of-flight (TOF) density distributions, allowing to distinguish phases A and B in experiment. Lower (upper) row:
$\braket{n\left(\vect{r}\right)}_{t_{\mathrm{tof}}}$ integrated along
$z$ ($x$) in arbitrary scale. 
For example, when viewed along the $z$-direction,  phase A displays a
doughnut form (lower  left panel)  because of an incoherent addition
of $p_x$- and $p_y$-Wannier envelopes.  
In phase B, the sites occupied by $\left(\ket{p_x}\pm i \ket{p_y}\right)/\sqrt{2}$ give a similar doughnut structure, but the hole at $k_x=k_y=0$ is filled by the other half of the sites with $p_z$-particles. 
Similarly, viewing along the $x$-direction reveals the existence of $p_z$-particles in phase B, contrary to phase A (upper row).
\label{fig:tofonethirdfillingV3is01}
}
\vspace*{-0.45cm}
\end{figure}

Experimentally, the different phases can, \textit{e.g.}, be
distinguished by measuring the density distribution after a time of
flight $t_{\mathrm{tof}}$. This relates to the in-trap momentum
distribution via 
$
	\braket{n\left(\vect{r}\right)}_{t_{\mathrm{tof}}}=
		\left[M/(\hbar\, t_{\mathrm{tof}})\right]^3 \sum_{\mu,\nu}w_{\mu}^\star\left(\vect{k}\right) w_{\nu}\left(\vect{k}\right)
		\braket{c_{\mu}^\dagger\left(\underline{\vect{k}}\right) c_{\nu}\left(\underline{\vect{k}}\right)}$,
with $w_{\mu}\left(\vect{k}\right)$ the
Fourier transform of the Wannier orbital
$w_{\mu}\left(\vect{r}\right)$, $c_{\mu}(\underline{\vect{k}})=\sum_{\vect{i}} \ue^{i
  \underline{\vect{k}} \cdot \vect{i}} c_{\mu,\vect{i}}/L^{3/2}$, and
$\vect{k}=M\,\vect{r}/(\hbar\, t_{\mathrm{tof}})$. $\underline{\vect{k}}$ is $\vect{k}$ modulo reciprocal lattice vectors. 
Features in the density distribution appear because of its non-trivial 
$p$-orbital Wannier envelope.
This allows to distinguish phases A and B by their column density
(\emph{i.e.}, the
density integrated along one spatial direction), 
see Fig.\ \ref{fig:tofonethirdfillingV3is01}.

Observation of these novel phases requires that we simultaneously achieve strong interactions, $V \gg t$ and low temperatures 
$k_B T \ll t^2/V$, for the characteristic tunneling rate $t$ and interaction energy $V$. At experimentally feasible temperatures, 
this requires a significant enhancement of the real part of the $p$-wave scattering volume via the OFR. In practice, however, this is 
limited by spontaneous emission, which broadens the resonance and also leads to recoil heating. For the example considered 
in~\cite{Goyal2010b} based on the $^1S_0$ $\rightarrow$ $^3P_1$ intercombination line in $^{171}$Yb, the atomic linewidth is $\approx$180 kHz, 
which limits the useful OFR $p$-wave enhancement. Other species such as $^{87}$Sr, where the same transition has a linewidth of 
$\approx$7.5 kHz, should result in a substantial OFR, with a reasonable linewidth. Experimental studies of OFRs in related isotopes are 
currently underway~\cite{Blatt2011}.

In summary, we investigated the orbital order of spinless fermions in the $p$-band of a cubic lattice with 
interaction controlled by an OFR. The system can be realized with current technology. 
The model Hamiltonian can be expressed elegantly by Gell-Mann matrices. 
We analyzed the orbital order in the strong-coupling limit at $p$-band filling 1/3 using a Gutzwiller-type ansatz. Besides a phase where all tunneling is blocked and an antiferro-orbital phase where $p_x$- and $p_y$-orbitals alternate, we found a novel phase with axial orbital order which not only breaks translational symmetry but also has macroscopic orbital angular momentum. 
We expect our results to stimulate future work on this subject. For example, it is interesting to investigate
how quantum fluctuations affect the phase diagram: they might distort it 
\cite{Toth2010} or even lead to disordered `orbital liquid' states.
Fluctuations are also expected to lift the degeneracy between $p_x$- and $p_y$ orbitals at $J_3=0$, and possibly lead to \emph{spontaneous} TRS breaking.
Moreover, phase $B_{\pm}$ may have interesting topological properties. For example, at an interface of two domains with $p_x+i p_y$ and $p_x-i p_y$ order, chiral zero mode fermions may arise. 
Finally, other lattices and the limit of small interactions, where related models show non-trivial color-superfluidity \cite{Rapp2007,Rapp2008,Miyatake2009}, are also interesting.

This work is supported in part by NSF Grant PHY05-51164. 
P.~H.\ and M.~L.\ acknowledge support by the
Caixa Manresa, 
Spanish MICINN 
(FIS2008-00784 
and Consolider QOIT), 
EU Projects AQUTE and NAMEQUAM, and
ERC Grant QUAGATUA. 
E.~Z.\ is supported by NIST Grant 70NANB7H6138 Am 001.
W.~V.~L.\ is supported by ARO (W911NF-11-1-0230) and DARPA-OLE
(ARO W911NF-07-1-0464).
K.~G.\ and I.~H.~D.\ are supported by NSF Grant PHY-0903953.

\bibliographystyle{aps}

\begin{thebibliography}{10}
\bibitem{footnoteCollisionsppsd}
For an empty $s$-band, energy conservation can suppress undesired collisions $p+p\to s+d$ when using anharmonic on-site potentials created by optical superlattices.
\bibitem{checkLeven} Further, we checked that for even $L$ the occurring phases do not depend on $L$.
\bibitem{footnoteOffsiteinteractions}
The derivation of \eqref{eq:Heffonethird} neglected repulsive off-site interactions $n_{\mu,\vect{i}}n_{\mu,\vect{i}+\vect{\delta}}$, 
which for $p$-band fermions might be important compared to the exchange couplings $J$. 
However, the alternating occupations in the (physically relevant) phases A and B also minimize possible energy contributions from off-site repulsions.
Hence, neglecting these is justified a posteriori.
\end{thebibliography}

\end{document}